\documentclass[journal]{IEEEtran}
\usepackage{filecontents}
\usepackage{caption}
\usepackage[noadjust]{cite}

\ifCLASSINFOpdf
\usepackage[pdftex]{graphicx}
\else
\fi
\usepackage{amsmath}
\usepackage{amssymb}
\usepackage{mathtools}
\DeclarePairedDelimiter\floor{\lfloor}{\rfloor}
\newcommand{\RNum}[1]{\uppercase\expandafter{\romannumeral #1\relax}}
\usepackage{adjustbox}
\usepackage{algorithmic}

\usepackage{array}
\usepackage{makecell}
\usepackage{multirow}
\newcolumntype{M}[1]{>{\centering\arraybackslash}m{#1}}
\newcolumntype{P}[1]{>{\centering\arraybackslash}p{#1}}

\begin{document}
%
\title{A New Signal Representation Using Complex Conjugate Pair Sums}
%
%
%

\author{Shaik~Basheeruddin~Shah,~\IEEEmembership{Student~Member,~IEEE,}
~Vijay~Kumar~Chakka,~\IEEEmembership{Senior~Member,~IEEE,}~and~Arikatla~Satyanarayana~Reddy 
\thanks{Shaik Basheeruddin Shah and Vijay Kumar Chakka are with the Department of Electrical Engineering, Shiv Nadar University, India (e-mail: bs600@snu.edu.in, Vijay.Chakka@snu.edu.in).}
\thanks{Arikatla Satyanarayana Reddy is with the Department of Mathematics, Shiv Nadar University, India (e-mail: satyanarayana.reddy@snu.edu.in).}}
\maketitle

\begin{abstract}
This letter introduces a real valued summation  known as {\em Complex Conjugate Pair Sum (CCPS)}. The space spanned by CCPS and its one circular downshift is called {\em Complex Conjugate Subspace (CCS)}. For a given positive integer $N\geq3$, there exists $\frac{\varphi(N)}{2}$ CCPSs forming $\frac{\varphi(N)}{2}$ CCSs, where $\varphi(N)$ is the {\em Euler's totient function}. We prove that  these CCSs are mutually orthogonal and their direct sum form a $\varphi(N)$ dimensional subspace $s_N$ of $\mathbb{C}^N$. We propose that any signal of finite length $N$ is represented as a linear combination of 
elements from a special basis of $s_d$, for each divisor $d$ of $N$.
This defines a new transform named as {\em Complex Conjugate Periodic Transform (CCPT)}.
Later, we compared CCPT with DFT (Discrete Fourier Transform) and RPT (Ramanujan Periodic Transform).
It is shown that, using CCPT we can estimate the period, hidden periods and frequency information of a signal.
Whereas, RPT  does not provide the frequency information. 
For a complex valued input signal, CCPT offers computational benefit over DFT.
A CCPT dictionary based method is proposed to extract non-divisor period information.
\end{abstract}

%
\IEEEpeerreviewmaketitle
\section{Introduction}
\IEEEPARstart{F}{inite} length signal representation is a fundamental concept in signal processing, which has been studied for years. There are some techniques like DFT, DCT (Discrete Cosine Transform) and RPT \cite{Oppenheim}, \cite{6839030}.
Here DFT and DCT are very well-known techniques, whereas RPT is newly introduced by P.P. Vaidyanathan, which is widely used in period estimation applications \cite{7527160},  \cite{7952298}, \cite{7838897}. Recently Srikanth V. Tenneti and P.P. Vaidyanathan introduced a specific kind of signal representation using Nested Periodic Matrices (NPMs) \cite{7109930}, which is useful for period estimation. RPT  and DFT are members of NPMs family. The signal representation proposed in this letter is inspired from NPMs and subspace-based signal representation \cite{4472242}, \cite{5419949}, \cite{7494930}. 
 
For a given positive integer $N\geq 3$, there exist $\frac{\varphi(N)}{2}$ pairs of complex conjugate exponential sequences, having period exactly equal to $N$ \cite{7544641}. By using each pair, a real valued summation known as complex conjugate pair sum is introduced.
Later, some of its properties like symmetricity, periodicity and orthogonality are studied. It is shown that, each proposed summation and its one circular downshift act as a basis for two-dimensional complex conjugate subspace. 
So, there are total $\frac{\varphi(N)}{2}$ complex conjugate pair sums leading to $\frac{\varphi(N)}{2}$ complex conjugate subspaces for a given positive integer $N\geq 3$. Utilizing the orthogonality property of complex conjugate pair sums, it is proved that any two complex conjugate subspaces are mutually orthogonal. 
Then the direct sum of all these $\frac{\varphi(N)}{2}$ complex conjugate subspaces forms a $\varphi(N)$ dimensional subspace $s_N$. 
We solve the problem of signal representation by considering the signals from divisor subspaces $s_d$, since $\sum_{{d}|N}^{}\varphi(d) = N$ \cite{Hardy}, \cite{6839014}.
Its corresponding transform is named as \textit{Complex Conjugate Periodic Transform}. 
It is a real transform like DCT, discrete sine transform, etc. Interestingly, it has NPM structure and estimates period information. 
Hence, complex conjugate periodic transform is compared with RPT and DFT in this letter.

%
%
%
%

The structure of the letter is as follows: Complex conjugate pair sums are introduced in Section \RNum{2}. A new basis for the complex conjugate subspace is presented in Section \RNum{3}. 
Finite length signal representation and its application are discussed in Section \RNum{4} and \RNum{5}.
Conclusions are drawn in Section \RNum{6}.
 
 \textit{Notation:}
Greatest common divisor (gcd) between two numbers $a$ and $b$ is denoted as $(a,b)$ and $lcm$ indicates least common multiple. 
For a given $n\in\mathbb{N}$, Euler's totient function $\varphi(n)$ is defined as $\varphi(n) = \#\{k\ |\ 1\ {\leq}\ k\ {\leq}\ n,\ (k,n)=1\}$. Since $(k,n) = (n-k,n)$, $\varphi(n)$ is even for $n\geq 3$. If $d$ is a divisor of $N$, then it is denoted as $d|N$. Symbol $\floor*{a}$, rounds the value $a$ to the greatest integer less than or equal to $a$. 
The conjugate transpose of a matrix $\mathbf{A}$ is denoted by $\mathbf{A^H}$.
\section{Complex Conjugate Pair Sum}
In DFT, an $N$-finite length signal is represented by using the following complex exponential basis,
\begin{equation}
\footnotesize
{s_{N,k}}(n) = e^{\frac{j2{\pi}kn}{N}},\quad 0\ {\leq}\ n,k\ {\leq}\ {N-1}.
\label{DFT_Basis}
\normalsize
\end{equation}
For a given $k$, we denote the period of ${s_{N,k}}(n)$ by $P_k$. It is known that $P_k = \frac{N}{(k,N)}.$ Hence $P_k|N.$
Let $D_N=\{d\in \mathbb{N}\mid d|N\}$ be the set of positive divisors of $N$. In order to count the number of $k\in \{0,1,\ldots,N-1\}$ having a particular period $d\in {D_N}$, it is sufficient to count the number of $k$ such that $(k,N)=\frac{N}{d},$ {\it i.e.,} $\# H_d$, where $H_d=\{k\in \mathbb{N}{\mid}0\le k\le N-1, (k,N)=\frac{N}{d}\}.$ Here it is easy to see that the sets $H_d$ for each $d\in D_N$ form a partition of $\{0,1,\ldots,N-1\}.$ And the set $H_d$  is equivalent to the set $\{x\in \mathbb{N}{\mid}0{\leq}x{\leq}{d-1},(x,d)=1\}.$ As a consequence, we have $\#H_d=\varphi(d)$ and $\sum_{d|N}\varphi(d) = N$.
From this discussion the following result is trivial.

\textit{Theorem 1:} The number of sequences in $\{s_{N,k}(n){\mid}$ $0{\leq}k{\leq}N-1\}$, having period exactly equal to $d$ is $\varphi(d)$, where $d|N$.

For every $s_{N,k}(n)$ there exists a complex conjugate sequence $s_{N,N-k}(n)$ as $(k,N) = (N-k,N)$.
These two sequences form a complex conjugate pair. So, for $N{\geq}3$, there are $\frac{\varphi(N)}{2}$ complex conjugate pair sequences, having period exactly equal to $N$. 
If $N=1$ (or) $2$, the sequence $s_{N,k}(n)$ itself is a complex conjugate,  because $\varphi(1) = \varphi(2) = 1$.
Given any $N\in\mathbb{Z^+}$, the Complex Conjugate Pair Sum (CCPS) is defined as,
\begin{equation}
\begin{aligned}
c_{N,k}(n) &= M\big(s_{N,k}(n)+s_{N,N-k}(n)\big)= 2Mcos\Big(\frac{2{\pi}{k}n}{N}\Big).
\label{CCPS_Def}
\end{aligned}
\end{equation}
where $\footnotesize M =\begin{cases}
	\frac{1}{2},& \text{if}\ N=1\ \text{(or)}\ 2\\
    1, & \text{if }\ {N{\geq}3}
\end{cases}\normalsize$, $k\in$ $A_N=\{a\in$ $\mathbb{N}{\mid}1{\leq}a{\leq}\floor*{\frac{N}{2}},$ $(a,N)=1\}$ if $N\ge 3$ and $k=1$ otherwise.
Notice that CCPSs are real, even symmetric \big($c_{N,k}(n) = c_{N,k}(N-n)$\big) and periodic with period $N$. 
Using the orthogonality between the complex exponentials, one can easily verify the following theorem.\\
\textit{Theorem 2: Orthogonality:} Any two CCPSs 
and their circular shifts are orthogonal to each other, {\it i.e.,}
\begin{equation}
\begin{aligned}
&\sum\limits_{n=0}^{N-1}c_{N_1,k_1}(n-{l_1})c_{N_2,k_2}(n-{l_2})\\& = 2N{M^2}cos\Bigg(\frac{2{\pi}{k_1}({l_1}-{l_2})}{N_1}\Bigg)\delta({N_1}-{N_2})\delta({k_1}-{k_2}),
\end{aligned}
\end{equation}
where $N = lcm(N_1,N_2)$, ${l_1}\in{\mathbb{Z}}$ and ${l_2}\in{\mathbb{Z}}$.
%
\section{A New Basis For Complex Conjugate Subspaces}
For each $k\in A_N$, we can construct an $N\times N$ circulant matrix $\mathbf{D_{N,k}}$ as given below:
\begin{equation}
\footnotesize
\mathbf{D_{N,k}} = \begin{bmatrix}
c_{N,k}(0) & c_{N,k}(N-1) & \dots & c_{N,k}(1) \\
c_{N,k}(1) & c_{N,k}(0) & \dots & c_{N,k}(2) \\
\vdots & \vdots & \ddots & \vdots\\
c_{N,k}(N-1) & c_{N,k}(N-2) & \dots & c_{N,k}(0)
\end{bmatrix}.
\normalsize
\end{equation}
Here each column in $\mathbf{D_{N,k}}$ is a circular downshift of the previous column. 
Using (\ref{CCPS_Def}), $\mathbf{D_{N,k}}$ can be written as
\footnotesize
\begin{equation}
\begin{aligned}
\mathbf{D_{N,k}} &= \begin{bmatrix}
e^{\frac{j2{\pi}k(0)}{N}}+e^{\frac{-j2{\pi}k(0)}{N}}&\dots&e^{\frac{j2{\pi}k(1)}{N}}+e^{\frac{-j2{\pi}k(1)}{N}} \\
e^{\frac{j2{\pi}k(1)}{N}}+e^{\frac{-j2{\pi}k(1)}{N}} & \dots & e^{\frac{j2{\pi}k(2)}{N}}+e^{\frac{-j2{\pi}k(2)}{N}} \\
\vdots & \ddots & \vdots\\
e^{\frac{j2{\pi}k(N-1)}{N}}+e^{\frac{-j2{\pi}k(N-1)}{N}} & \dots & e^{\frac{j2{\pi}k(0)}{N}}+e^{\frac{-j2{\pi}k(0)}{N}}
\end{bmatrix}\\&= \mathbf{B}\mathbf{B^H},
\end{aligned}
\label{Factorization}
\end{equation}
\normalsize
where 
\begin{equation}
\footnotesize
\mathbf{B^H}=\begin{bmatrix}
e^{\frac{-j2{\pi}{k}(0)}{N}}&e^{\frac{-j2{\pi}{k}(1)}{N}}&\dots&e^{\frac{-j2{\pi}{k}(N-1)}{N}} \\
e^{\frac{j2{\pi}{k}(0)}{N}}&e^{\frac{j2{\pi}{k}(1)}{N}}&\dots&e^{j\frac{2{\pi}{k}(N-1)}{N}}
\end{bmatrix}_{2\times N}.
\normalsize
\end{equation}
This property of decomposing $\mathbf{D_{N,k}}$ into $\mathbf{B}\mathbf{B^H}$ is known as \textit{factorization}. Since $rank(\mathbf{D_{N,k}})=rank(\mathbf{BB^H})=rank(\mathbf{B})$ \cite{Strang} and columns of $\mathbf{B}$ are orthogonal 
\cite{Oppenheim}, the $rank(\mathbf{D_{N,k}})$ is always equal to $2$, whatever the values of $N$ and $k$.
Further, it is easy to check that first two columns of $\mathbf{D_{N,k}}$ are linearly independent.

From (\ref{Factorization}), the column space of $\mathbf{D_{N,k}}$ (denoted as $v_{N,k}$) is the same as the column space of $\mathbf{B}$ \cite{Strang}, which consists of $p_i$ periodic signals having a particular frequency $\frac{2{\pi}k}{N}$.
In  \cite{7544641}, S. W. Deng et al., introduced a subspace, spanned by the columns of $\mathbf{B}$ known as \textit{Complex Conjugate Subspace (CCS)}. So, the subspace $v_{N,k}$ is also termed as CCS. This results in a new real basis for CCS. Any $x_{N,k}(n)\ {\in}\ v_{N,k}$ can be expressed as
\begin{equation}
\scriptsize
x_{N,k}(n) =\sum\limits_{l=0}^{1}{\beta_{lk}c_{N,k}(n-l)}.\normalsize
\label{CCS_Basis}
\end{equation}
Since $\#A_N = \frac{\varphi(N)}{2}$, we can construct $\frac{\varphi(N)}{2}$ circulant matrices, corresponding to the total $\frac{\varphi(N)}{2}$ CCSs. 
As CCPSs and its circular shifts are mutually orthogonal, whenever $N_1=N_2$ and $k_1\neq k_2$ (refer \textit{Theorem 2}), it follows that any two CCSs $v_{N,k_i}$ and $v_{N,k_j}$ are orthogonal, $\forall i\neq j$.
In summary, for a given $N\geq3$, there are $\frac{\varphi(N)}{2}$ two-dimensional orthogonal CCSs, each having CCPS and its one circular downshift as a basis. 
\section{Finite Length Signal Representation}
From the previous section, the direct sum of $\frac{\varphi(N)}{2}$ orthogonal CCSs form a subspace $s_N$ of dimension $\varphi(N)$ {\it i.e.,}
\begin{equation}
\footnotesize
s_N = v_{N,k_1}{\oplus}\ v_{N,k_2}\oplus\ \dots{\oplus}\ v_{N,k_{\frac{\varphi(N)}{2}}}.
\label{Directsum}
\normalsize
\end{equation}
Since $c_{N,k}(n)$ is an $N$ periodic sequence, any $x_N(n){\in}s_N$ is also an $N$ periodic sequence and it can be represented as,
\begin{equation}
\footnotesize
x_N(n)=\sum\limits_{\substack{{k}=1\\(k,N)=1}}^{\floor*{\frac{N}{2}}}x_{N,k}(n)= \sum\limits_{\substack{{k}=1\\(k,N)=1}}^{\floor*{\frac{N}{2}}}\sum\limits_{l=0}^{1}{\beta_{lk}c_{N,k}(n-l)},
\normalsize
\end{equation}
where $x_{N,k}(n){\in}v_{N,k}$ and $0{\leq}n{\leq}N-1$. 
Since $\sum_{d|N}\varphi(d) = N$, an $N$ length sequence $x(n)$ is represented as a linear combination of sequences $x_{d}\in s_{d}$, where $d\in D_N.$ Let $D_N=\{p_1,p_2,\ldots,p_m\}$, where $m=\#D_N,$ then
\begin{equation}
\footnotesize
x(n)=\sum_{{p_i}|N}x_{p_i}(n)=\sum_{{p_i}|N} \sum\limits_{\substack{{k}=1\\(k,p_i)=1}}^{\floor*{\frac{p_i}{2}}}\sum\limits_{l=0}^{1}{\beta_{l{k}i}c_{p_i,k}(n-l)},
\normalsize
\label{Synthesis}
\end{equation}
where $0{\leq}n{\leq}N-1$. Equation (\ref{Synthesis}) written in matrix form as,
\begin{equation}
\footnotesize
\mathbf{x} = \mathbf{T_N}\boldsymbol{\beta}=\begin{bmatrix}
\mathbf{R_{p_1}} & \mathbf{R_{p_2}} &\dots&\mathbf{R_{p_m}} 
\end{bmatrix}_{N{\times}N} \boldsymbol{\beta},
\label{Synthesis1}
\normalsize
\end{equation}
where $\boldsymbol{\beta}$ is a transform coefficient vector and $\mathbf{R_{p_i}}$ is the basis matrix for the subspace $s_{p_i}$, which is given below
\begin{equation}
\footnotesize
\mathbf{R_{p_i}}=\begin{bmatrix}
\hat{c}_{{p_i},{k_1}}&{\hat{c}^1}_{{p_i},{k_1}}&\dots&\hat{c}_{{p_i},{k_{\frac{\varphi(p_i)}{2}}}}&{\hat{c}^1}_{{p_i},{k_{\frac{\varphi(p_i)}{2}}}}
\end{bmatrix}_{N{\times}{\varphi(p_i)}},
\label{Synthesis2}
\normalsize
\end{equation}
where $\hat{c}_{{p_i},{k_i}}$ is an $N{\times}1$ sequence, obtained by repeating ${c}_{{p_i},{k_i}}$ periodically $\frac{N}{p_i}$ times and ${\hat{c}^1}_{{p_i},{k_i}}$ indicates one circular downshift of $\hat{c}_{{p_i},{k_i}}$. 
Since $c_{p_i,k_i}$ is not orthogonal to $c^1_{p_i,k_i}$ (refer \textit{Theorem 2}), the matrix $\mathbf{T_N}$ is not an orthogonal matrix. Now we show that $\mathbf{T_N}$ is an NPM. For the properties of NPM refer \cite{7109930} and \cite{7094814}.

\textit{Proposition 1:} For a given $N\in \mathbb{Z^+}$, $\mathbf{T_N}$ is an NPM.

\textit{Proof:} For each $i\in \{1,2,\dots,m\}$ the columns of $\mathbf{R_{p_i}}$ are periodic with period $p_i$ and form a basis for $s_{p_i}$, consequently the rank of $\mathbf{R_{p_i}}$ is $\varphi(p_i)$. From \textit{Theorem 2}, it is easy to verify that $\mathbf{R_{p_i}^T{R_{p_j}}} = 0,\ \forall\ {p_i}\neq{p_j}$. Hence the result follows by invoking $\sum_{p_i|N}^{}\varphi(p_i) = N$.

As mentioned above, $\mathbf{R_{p_i}^T{R_{p_j}}} = 0,$ $\forall\ {p_i}\neq{p_j}$, {\em i.e.,} the column space of $\mathbf{R_{p_i}}$ is orthogonal to column space of $\mathbf{R_{p_j}}$.
Whenever these subspaces are orthogonal, they can be uniquely determined as \textit{Ramanujan subspaces (RSs)} (refer \textit{Theorem 2} in \cite{7109930}). 
So the subspace defined in (\ref{Directsum}) is known as RS.
One of the important properties of RS, which is useful in period estimation is:
\begin{itemize}
\item Consider a $p-$periodic signal $\footnotesize x(n) = \sum_{i=1}^{m}x_{p_i}(n)\normalsize$, where the period of $x_{p_i}(n)$ is equal to $p_i$. In general $p|lcm(p_1,\dots,p_m)$, but if $x_{p_i}(n)\in s_{p_i}$, then $p=lcm(p_1,\dots,p_m)$ \cite{6839014}.
\end{itemize}

In the above theoretical framework, we considered $N{\geq}3$, 
if $N=1$ (or) $2$, the $rank(\mathbf{D_{N,k}})=1$, so, the one independent column is the sequence itself. In this case, the value of $l$ (in equation (\ref{CCS_Basis})) corresponding to circular downshift is chosen as $0$. An example of representing a $5$ length sequence $x(n)$ using (\ref{Synthesis1}) and (\ref{Synthesis2}), having divisors $(p_i)$ $1$ and $5$ is given below:
\begin{equation}
\scriptsize
\begin{aligned}
\mathbf{X} & = \underbrace{\begin{bmatrix}
{c_{1,1}}(0) & {c_{5,1}}(0) & {c_{5,1}}(4) & {c_{5,2}}(0) & {c_{5,2}}(4)\\
{c_{1,1}}(0) & {c_{5,1}}(1) & {c_{5,1}}(0) & {c_{5,2}}(1) & {c_{5,2}}(0)\\
{c_{1,1}}(0) & {c_{5,1}}(2) & {c_{5,1}}(1) & {c_{5,2}}(2) & {c_{5,2}}(1)\\
{c_{1,1}}(0) & {c_{5,1}}(3) & {c_{5,1}}(2) & {c_{5,2}}(3) & {c_{5,2}}(2)\\
{c_{1,1}}(0) & {c_{5,1}}(4) & {c_{5,1}}(3) & {c_{5,2}}(4) & {c_{5,2}}(3)
\end{bmatrix}}_{\mathbf{T_5}}
\underbrace{\begin{bmatrix}
\beta_{011}\\
\beta_{012}\\
\beta_{112}\\
\beta_{022}\\
\beta_{122}
\end{bmatrix}}_{\boldsymbol{\beta}}.
\end{aligned}
\normalsize
\end{equation}

The RS is spanned by Ramanujan sums and its circular shifts in Ramanujan Periodic Representation (RPR) \cite{6839030}. Whereas, RS is spanned by CCPSs and their one circular downshift in the proposed representation. 
Analogous to RPR the proposed representation is termed as \textbf{\textit{Complex Conjugate Periodic Representation (CCPR)}}.
The transformation from $\mathbf{x}$ to $\boldsymbol{\beta}$, {\it i.e.,} $\boldsymbol{\beta} = \mathbf{T_N^{-1}}\mathbf{x}$ is termed as \textbf{\textit{Complex Conjugate Periodic Transform (CCPT)}}.
\section{Application of CCPT}
\subsection{Period and Frequency Estimation Using CCPT}
In general, signals can be either real or complex valued. 
There are many fields like communication, electromagnetics, acoustics etc., where people work with complex valued signals \cite{5961645}, \cite{schreier2010statistical}. 
To extract the maximum benefit of the proposed concept, two complex valued sequences are considered as examples.

The first example is:
\begin{equation}
\footnotesize
\begin{aligned}
&y_1(n)= x_{11}(n)+x_{12}(n)+x_{13}(n)\\&
=e^{\left(j2{\pi}10\left(\frac{n}{360}\right)+\frac{\pi}{5}\right)}+e^{\left(j2{\pi}40\left(\frac{n}{360}\right)+\frac{\pi}{4}\right)}+e^{\left(j2{\pi}50\left(\frac{n}{360}\right)+\frac{\pi}{3}\right)},
\end{aligned}
\normalsize
\end{equation}
here $y_1(n)$ is a $72$ length periodic sequence with period $36$.
It is easy to see that the periods of $x_{11}(n)$, $x_{12}(n)$ and $x_{13}(n)$ are $36$, $9$ and $36$ respectively. 
That is, $y_1(n)$ is decomposed into signals whose periods known as \textit{hidden periods} which are less than or equal to the period of ${y}_1(n)$.

\begin{figure}[!h]
\centering
 \includegraphics[width=4.8in,height=0.9in]{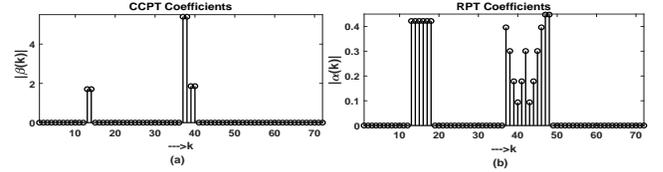} 
\caption[\footnotesize{(a)-(b) Absolute values of CCPT and RPT coefficients computed for $y_1(n)$ respectively.}]{\footnotesize{(a)-(b) Absolute values of CCPT and RPT coefficients computed for $y_1(n)$ respectively.}}
\label{f1}
\end{figure}

Fig.\ref{f1} (a)-(b) depicts the absolute values of transform coefficients obtained after applying CCPT and RPT techniques accordingly on $y_1(n)$.
Out of $72$ transform coefficients, the coefficient indices 
$13$ to $18$ belongs to $s_9$ and $37$ to $48$ belongs to $s_{36}$ respectively.
One can observe that the significant transform coefficient strength is present in both $s_9$ and $s_{36}$. This implies the period of $y_1(n)$ is equal to $lcm(9,36)$.
So, we can find out the period information in a signal by using the proposed transform and RPT.

Apart from period information, the proposed CCPT has an additional advantage over RPT. 
To understand this, 
consider the following question: is it possible to separate both $x_{11}(n)$ and $x_{13}(n)$ having same period $36$ with different frequencies, from CCPT and RPT coefficients belongs to $s_{36}$?
\begin{itemize}
\item  In RPT, $s_{36}$ is spanned by Ramanujan sum $c_{36}(n)$ and its circular shifts. Note that, $c_{36}(n)$ is generated by adding all $\varphi(36)$ complex exponentials, having period $36$ with different frequencies.
Due to this, 
it is not possible to separate any particular frequency component information from RPT coefficients. This is the reason
all $\varphi(36)$ coefficients corresponding to $s_{36}$ are non-zero in RPT.
\item \textit{Whereas in CCPT, $s_{36}$ is further decomposed as a direct sum of $\frac{\varphi(36)}{2}$ orthogonal CCSs} (refer equation (\ref{Directsum})).
These are $v_{36,1}$, $v_{36,5}$, $v_{36,7}$, $v_{36,11}$, $v_{36,13}$ and $v_{36,17}$, where each subspace consist of frequency $10Hz$, $50Hz$, $70Hz$, $110Hz$, $130Hz$ and $170Hz$ respectively. Since these subspaces are orthogonal, the information related to each frequency component is separable. 
As $y_1(n)$ has frequencies $10Hz$ and $50Hz$, the first four coefficients of $s_{36}$, corresponding to CCSs $v_{36,1}$ and $v_{36,5}$ are non-zero. 
\end{itemize}
So, it is possible to get the frequency information of a signal using CCPT, whereas it is not possible using RPT. 

The second example is:
\begin{equation}
y_2(n) = x_{21}(n)+x_{22}(n),\ 0{\leq}n{\leq}99,
\end{equation}
where, periods of $y_2(n)$, $x_{21}(n)$ and $x_{22}(n)$ are $35$, $5$ and $7$ respectively.
One period data of $x_{21}(n)$ and $x_{22}(n)$ follow Gaussian distribution with mean zero and variance one.

\begin{figure}[!h]
\centering
 \includegraphics[width=4.7in,height=0.8in]{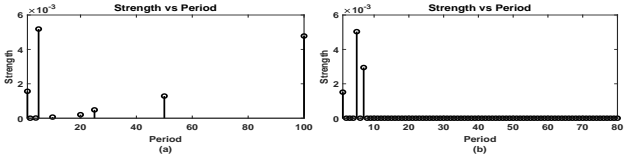} 
\caption[\footnotesize{(a) Strength of each divisor periodic component present in $y_2(n)$. (b) Strength of each periodic component present in $y_2(n)$.}]{\footnotesize{(a) Strength of each divisor periodic component present in $y_2(n)$. (b) Strength of each periodic component present in $y_2(n)$.}}
\label{f2}
\end{figure}

Now, CCPT coefficients are calculated for $y_2(n)$.
For each $p_i|100$, the absolute square sum of the $\varphi(p_i)$ CCPT coefficients corresponding to $s_{p_i}$, will give the strength of periodic component $p_i$ in the signal. 
Fig. \ref{f2}(a) depicts the strength of each $p_i$ present in $y_2(n)$ and it is easy to see that all $p_i$'s are having significant strength, because the period of $y_2(n)$ is not a divisor of the signal length.
Whereas period of $y_1(n)$ is a divisor of signal length in the previous example.
These results are obvious as we are considering only the divisor subspaces in finite length signal representation.   
Therefore, we are able to identify the divisor periods of a signal using CCPT.

As a summary, using CCPT, DFT (complex basis) and RPT (integer basis), we are able to find \textit{only} the divisor period information \cite{6839030}. TABLE \ref{tab:Comparison of different transformation techniques} summarizes the comparison between DFT, RPT and CCPT in different aspects.
\subsection{Non-Divisor Period Estimation}
In general, period and hidden periods of a signal are not divisors of signal length ($N$).
In literature \cite{6839030}, \cite{7109930}, \cite{7094814}, \cite{6853618}, there are two approaches used for both DFT and RPT to find out non-divisor periods.
Here we applied the same two approaches for CCPT and compared it with DFT and RPT.
\begin{table}[h]
\centering
\footnotesize
\caption{C\scriptsize{OMPARISON OF DIFFERENT TRANSFORMATION TECHNIQUES}}
\label{tab:Comparison of different transformation techniques}
\begin{adjustbox}{max width=\textwidth}
\renewcommand{\arraystretch}{1.25}
\scalebox{0.7}{
\begin{tabular}{|M{0.8cm}|M{1cm}|M{1.8cm}|M{1.5cm}|M{2.1cm}|}\hline 
\bfseries{\small{}}	&	\bfseries{\small{Divisor Period}} & \bfseries{\small{Non-divisor Period}}	&	\bfseries{\small{Frequency}} & \bfseries{\small{Complexity}}\\    \hline
\small{DFT}	&	\checkmark	& 	$\mathbb{\times}$	&	\checkmark & 	\small{$4N^2$-Real multiplications}	\\	\hline
\small{RPT}	&	\checkmark	& 	$\mathbb{\times}$	&	$\mathbb{\times}$ & 	\small{$2N^2$-Real multiplications}	\\	\hline
\small{CCPT}	&	\checkmark	& 	$\mathbb{\times}$	&	\checkmark & 	\small{$2N^2$-Real multiplications}	\\	\hline
\end{tabular}}
\end{adjustbox}
\normalsize
\end{table}
\subsubsection{First Approach} As proposed in \cite{6839030}, the transform coefficients are computed for the range of lengths $[N_1,N]$, where $N_1 < N$. 
In this way the signal is projected in every divisor subspace of $N_i$, where $N_1{\leq}N_i{\leq}N$. Then, by comparing the projection strength in each subspace we can find out the hidden periods in a
signal. 

Now, this method is applied on $y_2(n)$ by considering the range of $N_i$ as $[70,100]$. TABLE \ref{Significant hidden periods present in $y_2(n)$ for different signal length ($N_i$) values)} shows the hidden periods having significant strength in $y_2(n)$, for few $N_i$ values. One important observation of the results
is that, if any one of the hidden periods is a divisor of $N_i$, then the significant period strength exists in both $N_i$ and hidden period. 
Otherwise, the period strength is distributed on most of the divisors of $N_i$. 
From TABLE \ref{Significant hidden periods present in $y_2(n)$ for different signal length ($N_i$) values)}, it is also evident that along with hidden periods, it gives spurious hidden periods.
There are two more drawbacks to this approach along with this limitation.
\begin{itemize}
\item There is a huge possibility of divisors overlaps. 
This leads to the projection strength computation of a particular subspace multiple times.
\item \textit{Computational Complexity:}
Using CCPT, DFT and RPT basis requires
$\scriptsize 2\left(\frac{N^3-N_1^3}{3}+\frac{N^2+N_1^2}{2}+\frac{N-{N_1}}{6}\right)\normalsize$ number of real, complex and real multiplications respectively. 
\end{itemize}
Since $N_1{\leq}N_i{\leq}N$, the radix-$2$ FFT algorithm does not help in reducing the computational complexity of DFT.
\begin{table}[h]
\centering
\footnotesize
\caption{S\scriptsize{IGNIFICANT HIDDEN PERIODS PRESENT IN $y_2(n)$ FOR DIFFERENT SIGNAL LENGTH ($N_i$) VALUES}}
\label{Significant hidden periods present in $y_2(n)$ for different signal length ($N_i$) values)}
\begin{adjustbox}{max width=\textwidth}
\renewcommand{\arraystretch}{1.25}
\scalebox{0.7}{
\begin{tabular}{|M{1cm}|M{1.4cm}|M{1cm}|M{1.4cm}|M{1cm}|M{1.4cm}|}\hline 
\bfseries{\small{$N_i$}}	&	\bfseries{\small{Hidden periods}} & \bfseries{\small{$N_i$}}	&	\bfseries{\small{Hidden periods}} & \bfseries{\small{$N_i$}}	&	\bfseries{\small{Hidden periods}} \\    \hline
\small{70}	&	\small{5,7}	& 	\small{82}	&	\small{41,82} & 	\small{94}	&	\small{47,94}	\\	\hline
\small{72}	&	\small{24,36,72}	& 	\small{84}	&	\small{7,42} & 	\small{95}	&	\small{5,95}	\\	\hline
\small{76}	&	\small{19,38,76}	& 	\small{85}	&	\small{5,85} & 	\small{96}	&	\small{32,48,96}	\\	\hline
\small{77}	&	\small{7,77}	& 	\small{88}	&	\small{44,88} & \small{98}	&	\small{7,98}	\\	\hline
\end{tabular}}
\end{adjustbox}
\normalsize
\end{table}
Due to these drawbacks, the following method is suggested to estimate non-divisor periods.
\subsubsection{Second Approach}
In this approach, a signal $\mathbf{x}$ is projected into each and every subspace from $s_1$ to ${s_p}_{max}$, where $p_{max}$ is the maximum possible hidden period exists in $\mathbf{x}$.
This leads to define the synthesis dictionary model \cite{7094814}, \cite{6853618} as,

\qquad $\ \footnotesize \mathbf{x} = \mathbf{A}\mathbf{b} = [\mathbf{R_1},\mathbf{R_2},\dots,\mathbf{{R_p}_{max}}]_{N\times \hat{N}}\mathbf{b}_{\hat{N}\times 1},\normalsize$ \\
where $\hat{N} = \sum\limits_{i=1}^{p_{max}}\varphi(i)$, $\mathbf{b}$ is the transform coefficient vector and $\mathbf{A}$ is known as \textit{CCPT dictionary (fat matrix)}. 
Since $\hat{N}>>N$, there exist multiple solutions for $\mathbf{b}$. 
Now, considering the computational time as the main criteria, the non-divisor period estimation using dictionary is formulated as a \textit{data fitting} problem.
To get the best fit of the given signal with the signals having smaller periods, an optimization problem is formulated as follows \cite{7094814}, \cite{6853618}:
$\footnotesize min\ ||\mathbf{D}\mathbf{b}||_2\quad\text{s.t.}\quad\mathbf{x} = \mathbf{A}\mathbf{b}.\normalsize$
If $p_i$ is the period of $i^{th}$ column in $\mathbf{A}$ and $f(p_i) = {p_i}^2$, then the diagonal matrix $\mathbf{D}$ (known as penalty matrix) consist of $f(p_i)$ as $i^{th}$ diagonal entry. Now finding the optimal solution $(\mathbf{\hat{b}})$ to the above problem leads to the following closed form expression:

\qquad\qquad\quad $\footnotesize
\mathbf{\hat{b}} = \mathbf{{D^{-2}}}{\mathbf{A^T}}(\mathbf{A}\mathbf{D^{-2}}\mathbf{A^T})^{-1}\mathbf{x}.
\normalsize $\\
The $\varphi(p_i)$ coefficients in $\mathbf{\hat{b}}$, gives the strength of period $p_i$ in the signal. Fig. \ref{f2}(b) shows the hidden periods present in $y_2(n)$ by using the synthesis CCPT dictionary model with $p_{max} = 80$.
These are $1$, $5$, $7$ and period of the signal is $lcm(1,5,7)$.
Refer \cite{7109930}, \cite{7094814} and \cite{6853618} for the detailed theory of dictionary based method and for the result of synthesis dictionary model using DFT (Farey) and RPT dictionaries.

So, the proposed CCPT dictionary can be used to estimate non-divisor period and hidden periods of a signal.
Moreover, if the given signal consists a particular frequency, then we can estimate it from $\mathbf{\hat{b}}$, computed by using CCPT and Farey dictionaries, but it is not possible by using RPT dictionary.

TABLE \ref{tab:Comparison of different dictionary techniques} summarizes the above discussion including complexity, where $L$ used in TABLE \ref{tab:Comparison of different dictionary techniques} indicates the number of real multiplications required using CCPT dictionary.
\begin{table}[h]
\centering
\footnotesize
\caption{C\scriptsize{OMPARISON OF DIFFERENT DICTIONARY TECHNIQUES}}
\label{tab:Comparison of different dictionary techniques}
\begin{adjustbox}{max width=\textwidth}
\renewcommand{\arraystretch}{1.25}
\scalebox{0.7}{
\begin{tabular}{|M{1.6cm}|M{0.9cm}|M{1.8cm}|M{1.5cm}|M{2cm}|}\hline 
\bfseries{\small{}}	&	\bfseries{\small{Divisor Period}} & \bfseries{\small{Non-divisor Period}}	&	\bfseries{\small{Frequency}} & \bfseries{\small{Complexity}}\\    \hline
\small{Farey dictionary}	&	\checkmark	& 	\checkmark	&	\checkmark & 	\small{$2L$-Real multiplications}	\\	\hline
\small{RPT dictionary}	&	\checkmark	& 	\checkmark	&	$\mathbb{\times}$ & 	\small{$L$-Real multiplications}	\\	\hline
\small{CCPT dictionary}	&	\checkmark	& 	\checkmark	&	\checkmark & 	\small{$L$-Real multiplications}	\\	\hline
\end{tabular}}
\end{adjustbox}
\normalsize
\end{table}
If the given signal is real valued, then the complexity of both CCPT and DFT is almost same due to the conjugate symmetry property of DFT basis/dictionary coefficients.

\section{Conclusion}
In this work, the problem of a finite length signal representation is solved by using the signals belongs to CCSs. 
Later, we have shown that the proposed transformation (CCPT) allows us to get both divisor and non-divisor period information of a signal.
We compared the proposed method with DFT and RPT for a given complex (or) real valued input sequence.
Although this letter introduces CCPT, 
there are many other properties that require a detailed analysis. 
We will aspire to work on this in the near future.
\section*{Acknowledgement}
Authors would like to thank Dr.Krishnan Rajkumar and the four anonymous reviewers, whose comments are helpful to improve the letter quality.

\ifCLASSOPTIONcaptionsoff
  \newpage
\fi



\bibliographystyle{ieeetr}
\bibliography{bibs_1}
\end{document}